\documentclass[a4paper]{jpconf}
\usepackage{graphicx}
\usepackage{epstopdf}
\usepackage{hyperref}
\usepackage{bm}
\usepackage{amsmath}
\usepackage{amssymb}
\usepackage{times}
\usepackage{mathtools}
\usepackage{cite}
\begin{document}

\title{Structural properties of fractals with positive Lebesgue measures}

\author{E M Anitas$^{1,2}$}

\address{$^{1}$Joint Institute for Nuclear Research, Dubna 141980, Moscow region, Russia}
\address{$^{2}$Horia Hulubei National Institute of Physics and Nuclear Engineering, RO-077125 Bucharest, Romania}

\ead{anitas<at>theor.jinr.ru}

\begin{abstract}
Small-angle scattering (SAS) data which show a succession of power-law decays with decreasing values of scattering exponents, can be described in terms of fractal structures with positive Lebesgue measure (fat fractals). In this work we present a theoretical model for fat fractals and show how one can extract structural information about the underlying fractal using SAS method, for the well known fractals existing in the literature: Vicsek and Menger sponge. We calculate analytically the fractal structure factor and study its properties in momentum space. The models allow us to obtain the fractal dimension at each structural level inside the fractal, the number of particles inside the fractal and about the most common distances between the center of mass of the particles.
\end{abstract}

\section{Introduction}

An important technique for investigating the structure of various types of materials is small-angle scattering (SAS)~\cite{glatter82:book,svergun87:book} which yields the differential elastic cross section per unit solid angle as a function of the momentum transfer. It is known that many disordered systems show the property of self-similarity across nano- and microscales~\cite{gouyet96:book}, such as various types of elastomeric membranes~\cite{balasoiu08,anitas09}, cements~\cite{dasJACR14}, semiconductors~\cite{choJIEC14}, magnetic~\cite{anitasJAC14,naitoEPJB14} or biological structures~\cite{yadav14,gebhardtJACR14,majiPRE14}, and therefore the concept of fractal geometry~\cite{mandelbrot83:book,gouyet96:book} coupled with elements of SAS technique can give new insights regarding the structural characteristics of such fractal systems~\cite{schmidt86,martin87,schmidt91,chernyJACR10,chernyJSI,chernyPRE11,chernyJACR14}. One of the main parameters which can be obtained is the fractal dimension $D$~\cite{mandelbrot83:book,gouyet96:book}. For a mass fractal it is given by the scattering exponent of the power-law SAS intensity $I(q)~\propto~q^{-D}$ where $0<\mathrm{D}<3$.

However, a number of experimental SAS data shows a succession of power-laws whose scattering exponents take arbitrarily decreasing values~\cite{zhao09,headen09,golosova12}. Such a behavior was not clearly understood until recently~\cite{anitasEPJB14}, when it was shown that such type of successions corresponds to a fat fractal structure. In the later work, the main structural characteristics have been obtained by calculating the mono- and polydisperse structure and form factor for a system consisting of fat Cantor fractals (known also as $\epsilon$-Cantor sets~\cite{aliprantis98:book}). Although the fattened version of regular Cantor set (known also as thin Cantor sets) provides a simple and intuitive picture of the investigated properties, real physical systems have a more complex structure whose properties can be investigated by considering more complex structural models. 
 
The aim of this paper is to extend the fat Cantor fractal model to fractal structures which can resemble more closely real physical systems and to show how we can extract information about the fractal dimensions at each level, total number of particles and about the most common distances in the fractal. To this end, we calculate the structure factor of fattened versions of Vicsek~\cite{vicsek89:book} and Menger sponge~\cite{gouyet96:book}, as they are often considered as structural models for various natural or artificial systems. 

\section{Fat fractals}

An intuitive way to understand fat fractals is the following: we start with an initial cube (here is used a cube, but any other shape could be used as well; $m=0$, $m$ being the fractal iteration number) which is divided into 27 smaller cubes with side length $1/3$ from the initial cube. Generally, in the first iteration ($m=1$) some cubes (out of the 27 smaller ones) are kept and the others are removed. In particular, for the Vicsek fractal we keep the eight cubes in the corners plus the middle one and remove all the others, while for the Menger sponge we remove the smaller cubes in the middle of each face together with the cube in the center of the larger cube, and keep all the others. We repeat the same operation on each of the remaining cubes (9 for Vicsek, and respectively 20 for Menger), thus leaving 81 cubes, and respectively 400 cubes, of side length $1/3^{2}$ ($m=2$) and so on. Therefore, at the $m$-th iteration the side length of the remaining cubes is $1/3^{m}$. The number of remaining cubes then is given by
\begin{equation}
\label{number:eq}
N_{m} = \begin{dcases}
   9^{m},& for~Vicsek~fractals,\\
   20^{m}, & for~Menger~fractals,
  \end{dcases}
\end{equation}
The thin version  of these fractals is obtained in the limit $m \rightarrow \infty$, and have the same fractal dimensions $\log{9}/\log{3}=2$ (for Vicsek) and respectively $\log{20}/\log{3}=2.73$ (for Menger), independent on the iteration number. In this paper, the "fattened" version of these thin fractals is obtained by keeping the cubes instead of side length $1/3$ (for $m=1,2,3$), then $1/3^{2}$ (for $m=4,5,6$), then $1/3^{3}$ (for $m=7,8,9$), etc. The resulting fractal is topologically equivalent to the thin version, but the holes decrease in size sufficiently fast so that, when $m \rightarrow \infty$, the fractal has nonzero and finite volume, and fractal dimension 3. Here, in the construction process we consider the same rules as already described in~\cite{anitasEPJB14}: the side length of the initial cube is $l_0$ (called zero-order iteration or initiator) and specify it in Cartesian coordinates as a set of points satisfying the conditions $-l_{0}/2 \leq x \leq l_{0}/2$, $-l_{0}/2 \leq y \leq l_{0}/2$, $-l_{0}/2 \leq z \leq l_{0}/2$. The origin lies in the cube center, and the axes are parallel to the cube edges. 

The iteration rule (\textit{generator}) is to replace the initial cube by cubes of edge $\beta_{s}^{(1)}l_{0}$ ($m=1$; whenever the quantity $(...)$ appears in the exponent, it is to be interpreted as an index and not a power). For the Vicsek fractal the position of one cube is at the origin and the center of the eight cubes are shifted from the origin by the vectors $\bm{a}_{j}=\{\pm \beta_{t}^{(1)}l_{0}, \pm \beta_{t}^{(1)}l_{0}, \pm \beta_{t}^{(1)}l_{0}\}$ with all the combinations of the signs, where $\beta_{t}^{(1)} \equiv (1-\beta_{\mathrm{s}}^{(1)})/2$ and $\beta_{\mathrm{s}}^{(1)}$ is a dimensionless positive parameter for the first iteration, obeying the condition 
$0 < \beta_{\mathrm{s}}^{(1)} < 1/3$, in order to avoid overlapping of the cubes. For the Menger sponge the position of the eight cubes are shifted from the origin by the vectors $\bm{a}_{j}=\{\pm \beta_{t}^{(1)}l_{0}, \pm \beta_{t}^{(1)}l_{0}, \pm \beta_{t}^{(1)}l_{0}\}$ and the positions of the 12 cubes are shifted by the vectors $\{\pm \beta_{t}^{(1)}l_{0}, \pm \beta_{t}^{(1)}l_{0}, 0\}$, $\{\pm \beta_{t}^{(1)}l_{0}, 0, \pm \beta_{t}^{(1)}l_{0}\}$ and $\{0, \pm \beta_{t}^{(1)}l_{0}, \pm \beta_{t}^{(1)}l_{0}\}$ with all the combinations of the signs. The second and third iterations ($m=2,3$) are obtained by performing an analogous operation to each cube of the first iteration and with the same scaling factor $\beta_{\mathrm{s}}^{(1)}$. For each subsequent iterations we repeat the same operation but for $m=4,5$ and $m=6$ we take the scaling factor $\beta_{\mathrm{s}}^{(2)}$, for $m=7,8$ and $m=9$ we take the scaling factor $\beta_{\mathrm{s}}^{(3)}$ and so on. If one consider that the edge of the removed parallelepiped at iteration $m$ is $\gamma_{m}=\alpha^{p_{m}}$
where $1/3 < \alpha < 1$ and the exponent $p_{m}$ is defined as
\begin{equation}
p_{m} \equiv \begin{dcases}
   1, & m = 1, 2, 3\\
   \cdots \\
   k, & m = 3k-2, 3k,
\end{dcases}
\label{eq:exponent}
\end{equation}
where $m=1,2,\cdots$, then $\beta_{\mathrm{s}}^{(m)}=(1-\gamma_{m})/2$, the side length of each cube by $l_{m}=l_{0}/2^{m}\prod_{i=1}^{m}(1-\gamma_{i})$ and the components of the $\bm{a}_{j}$ vectors are given by
$\beta_{t}^{(m)}=l_{m-1}(1+\gamma_{m})/4$. Here, we consider $\alpha = 0.7$ in order to avoid superposition of cubes at least up to $m=6$.

From their construction one can see that the fat fractals are built from a succession of exact self-similar fractals having different scaling factors at different scales. Since the scaling factor depends on the iteration, each scale will have a different fractal dimension given by~\cite{chernyJACR10}
$D_\mathrm{m}=-3 \ln 2/\ln \beta_{s}^{(m)}$, and each of this scale gives a power-law decay of structure factor $S(q)~\propto~q^{-D_{\mathrm{m}}}$~\cite{chernyJACR10,chernyPRE11}.

\section{Results and discussions}

\begin{figure}
\begin{center}
\includegraphics[width=\columnwidth]{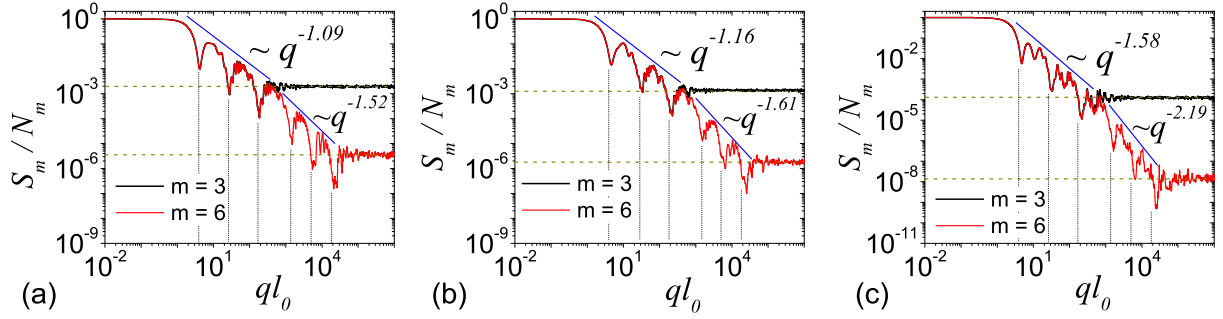}
\end{center}
\caption{
Monodisperse fractal structure factor (Eq.~(\ref{eq:sffinal})) for the third and sixth iterations for a) Cantor (included for comparison, see details in~\cite{anitasEPJB14}), b) Vicsek and c) Menger sponge, in units of $N_{m}$ (Eq.~(\ref{number:eq})), at $\alpha = 0.7$. The position of minima indicated by vertical dotted-lines are estimated from Eq.~(\ref{eq:minimaconditions})). The horizontal lines represent the asymptotes of the structure factor.}
\label{fig:fig1}
\end{figure}

\begin{figure}
\begin{center}
\includegraphics[width=0.99\columnwidth]{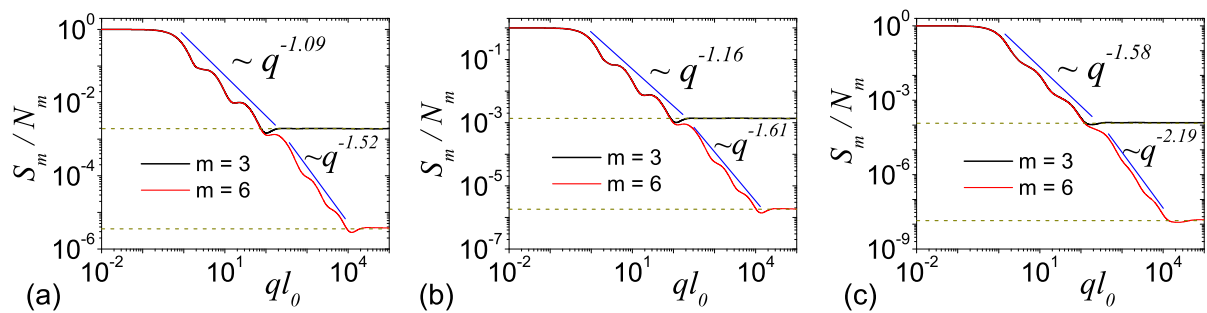}
\end{center}
\caption{
Polydisperse fractal structure factor (Eq.~(\ref{eq:sffinal}) together with the log-normal distribution~\cite{chernyPRE11}) for the third and sixth iterations for a) Cantor (included for comparison, see details in~\cite{anitasEPJB14}), b) Vicsek and c) Menger sponge, in units of $N_{m}$ (Eq.~(\ref{number:eq})), at $\alpha = 0.7$ and relative variance $\sigma_{\mathrm{r}}=0.4$. The horizontal lines represent the asymptotes of the structure factor.}
\label{fig:fig2}
\end{figure}

For structures corresponding to a generalized Vicsek fractal~\cite{chernyPRE11} and Menger sponge, the generative function, which gives the positions of the centers of scattering cubes at each iteration, reads as
\begin{equation}
G_{i}(\bm{q}) = \begin{dcases}
   \frac{1}{9} \left(1+C_{i}(\bm{q})\right), & for~Vicsek~fractals,\\
   \frac{1}{20}\left ( C_{i}(\bm{q})+4(L_{i}^{xz}(\bm{q})+L_{i}^{xy}(\bm{q})+L_{i}^{yz}(\bm{q}) \right ), & for~Menger~fractals,
\end{dcases}
\label{eq:gf}
\end{equation}
where $G_{0}(\bm{q}) \equiv 1$, $C_{i}(\bm{q})=8\cos(q_{x}u_{i})\cos(q_{y}u_{i})\cos(q_{z}u_{i}))$, $L_{i}^{xy}(\bm{q})=\cos(q_{x}u_{i})\cos(q_{y}u_{i})$, $L_{i}^{xz}(\bm{q})=\cos(q_{x}u_{i})\cos(q_{z}u_{i})$, $L_{i}^{yz}(\bm{q})=\cos(q_{y}u_{i})\cos(q_{z}u_{i})$ and $u_{i}=l_{0}\beta_{\mathrm{t}}^{(i)}\prod_{j=1}^{i-1}\beta_{\mathrm{s}}^{(j)}$.
Then, the structure factor is obtained (Figs.~(\ref{fig:fig1}) and~(\ref{fig:fig2})) from~\cite{chernyPRE11}
\begin{equation}
S(q)=N_{m} \left \langle  \prod_{i=1}^{m}|G_{i}(\bm{q}u_{i})|^{2} \right \rangle.
\label{eq:sffinal}
\end{equation}
The position of minima are obtained when the cubes inside the fractal interfere out of phase, and therefore using $2u_{m}=\pi/q$, for $k=1,\cdots,m$ we obtain (vertical lines in Fig.~(\ref{fig:fig1})
\begin{equation}
q_{k}l_{0}\simeq \frac{\pi}{2\beta_{\mathrm{t}}^{(k)}\prod_{i=1}^{k}{\beta_{\mathrm{s}}^{(i)}}}.
\label{eq:minimaconditions}
\end{equation}

A more accurate description of a real physical system is obtained by considering an ensemble of fractals with different sizes taken at random and distributed according to a log-normal distribution function with relative variance $\sigma_{\mathrm{r}}$~\cite{chernyJACR10}. The results are shown in Fig.~(\ref{fig:fig2}), where $\sigma_{\mathrm{r}}=0.4$. 

One can be clearly observe that beyond the last power-law decay (or generalized power-law decay in the monodisperse case) we have $q \gtrsim 1/u_{m}$ and therefore, in this region the fractal structure factor given by Eq.~(\ref{eq:sffinal}) is $S(q)\simeq 1$ and the asymptotic values are $1/N_{m}$ (Figs.~(\ref{fig:fig1}) and~\ref{fig:fig2})), as for the case of thin fractals~\cite{chernyJACR10,chernyPRE11}.

\section{Conclusions}
We have extended the fat Cantor fractal model to other well known types of fractal structures described in literature (Vicsek and Menger sponge) which can model more closely real physical systems, and have presented an analytically expression for the fractal structure factor in mono- and polydisperse case. We have obtained the fractal dimensions at each level (from  the slope of each power-law decay), total number of particles (from the asymptote of the structure factor) and the minima positions in the monodisperse factor (using Eq.~\ref{eq:minimaconditions}). The model can be used to describe the structure of nano- and micro clusters and it can be applied to experimental SAS data showing a succession of power-law decay with decreasing values of scattering exponents ("concave"-like scattering curves).

\section*{References}
\providecommand{\newblock}{}


\begin{thebibliography}{10}
\expandafter\ifx\csname url\endcsname\relax
  \def\url#1{{\tt #1}}\fi
\expandafter\ifx\csname urlprefix\endcsname\relax\def\urlprefix{URL }\fi
\providecommand{\eprint}[2][]{\url{#2}}

\bibitem{glatter82:book}
Glatter O and Kratky O 1982 {\em Small-angle X-ray Scattering\/} (London:
  Academic Press)

\bibitem{svergun87:book}
Feigin L~A and Svergun D~I 1987 {\em Structure Analysis by Small-Angle X-Ray
  and Neutron Scattering\/} (NY: Plenum press)

\bibitem{gouyet96:book}
Gouyet J~F 1996 {\em Physics and Fractal Structures\/} (Berlin: Springer)

\bibitem{balasoiu08}
Balasoiu M et al 2008 {\em Optoelectronics and Advanced Materials - Rapid
  Communications\/} {\bf 2} 730--734

\bibitem{anitas09}
Anitas E~M, Balasoiu M, Bica I, Osipov V~A and Kuklin A~I 2009 {\em
  Optoelectronics and Advanced Materials - Rapid Communications\/} {\bf 3}
  621--624

\bibitem{dasJACR14}
Das A, Mazumder S, Sen D, Yalmali V, Shah J~G, Ghosh A, Sahu A~K and Wattal P~K
  2014 {\em J. Appl. Cryst.\/} {\bf 47} 421--429

\bibitem{choJIEC14}
Cho K, Biswas P and Fraundorf P 2014 {\em J. Ind. Eng. Chem.\/} {\bf 20} 558

\bibitem{anitasJAC14}
Craus M~L, Islamov A~K, Anitas E~M, Cornei N and Luca D 2014 {\em Journal of
  Alloys and Compounds\/} {\bf 592} 121--126

\bibitem{naitoEPJB14}
Naito T et al 2014 {\em Eur. Phys. J. B\/} {\bf 86} 410

\bibitem{yadav14}
Yadav I, Kumar S, Aswal V~K and Kohlbrecher J 2014 {\em Phys. Rev. E\/} {\bf
  89} 032304--1

\bibitem{gebhardtJACR14}
Gebhardt R 2014 {\em J. Appl. Cryst.\/} {\bf 47} 29--34

\bibitem{majiPRE14}
Maji J, Bhattacharjee S~M, Seno F and Trovato A 2014 {\em Phys. Rev. E\/} {\bf
  89} 012121

\bibitem{mandelbrot83:book}
Mandelbrot B 1983 {\em The Fractal Geometry of Nature\/} (USA: W.H. Freeman)

\bibitem{schmidt86}
Schmidt P~W and Dacai X 1986 {\em Phys. Rev. A\/} {\bf 33} 560--566

\bibitem{martin87}
Martin J~E and Hurd A~J 1987 {\em J. of Appl. Cryst.\/} {\bf 20} 61--78

\bibitem{schmidt91}
Schmidt P~W 1991 {\em J. of Appl. Cryst.\/} {\bf 24} 414--435

\bibitem{chernyJACR10}
Cherny A~Y, Anitas E~M, Kuklin A~I, Balasoiu M and Osipov V~A 2010 {\em J.
  Appl. Cryst.\/} {\bf 43} 790--797

\bibitem{chernyJSI}
Cherny A~Y, Anitas E~M, Kuklin A~I, Balasoiu M and Osipov V~A 2010 {\em J.
  Surf. Invest.\/} {\bf 4} 903--907

\bibitem{chernyPRE11}
Cherny A~Y, Anitas E~M, Kuklin A~I and Osipov V~A 2011 {\em Phys. Rev. E\/}
  {\bf 84} 036203--1--036203--11

\bibitem{chernyJACR14}
Cherny A~Y, Anitas E~M, Osipov V~A and Kuklin A~I 2014 {\em J. Appl. Cryst.\/}
  {\bf 47} 198--206

\bibitem{zhao09}
Zhao J, Shi D and Lian J 2009 {\em Carbon\/} {\bf 47} 2329--2336

\bibitem{headen09}
Headen T~F, Boek E~S, Stellbrink J and M S~U 2009 {\em Langmuir\/} {\bf 25}
  422--428

\bibitem{golosova12}
Golosova A~A et al, 2012 {\em J. Phys. Chem. C\/} {\bf 116} 15765--15774

\bibitem{anitasEPJB14}
Anitas E~M 2014 {\em Eur. Phys. J. B\/} {\bf 87} 139

\bibitem{aliprantis98:book}
Aliprantis C~D and Burkinshaw O 1998 {\em Principles of Real Analysis\/}
  (Academic press)

\bibitem{vicsek89:book}
Vicsek T 1989 {\em Fractal Growth Phenomena\/} (Singapore: World Scientific)

\end{thebibliography}
\end{document}